\documentstyle[preprint,aps]{revtex}

\tightenlines
\input{epsfig.sty}
\newcommand{\Slash}[1]{\ooalign{\hfil/\hfil\crcr$#1$}}
\begin{document}

\preprint{}


\title{Dirac Sea Effects on Superfluidity in Nuclear Matter}

\author{Masayuki Matsuzaki
 \thanks{Electronic address: matsuza@fukuoka-edu.ac.jp}}

\address{Department of Physics, Fukuoka University of Education, 
Munakata, Fukuoka 811-4192, Japan}
\date{\today}
\maketitle

\begin{abstract}
We study two kinds of Dirac sea effects on the $^1S_0$ pairing gap in 
nuclear matter based on the relativistic Hartree approximation to quantum 
hadrodynamics and the Gor'kov formalism. We show that the vacuum fluctuation 
effect on the nucleon effective mass is more important than the direct 
coupling between the Fermi sea and the Dirac sea due to the pairing 
interaction. The effects of the high-momentum cutoff are also discussed.
\end{abstract}


 Superfluidity caused by the pairing correlation between two nucleons with 
the linear or angular momenta opposite to each other is a key ingredient to 
describe quantitatively the thermal evolution of neutron stars and the 
structure of finite nuclei. As a way of description, relativistic models 
are attracting attention. The origin of relativistic nuclear models can be 
traced back to the work of Duerr~\cite{duerr}. Since Chin and Walecka 
succeeded in reproducing the saturation property of symmetric nuclear matter 
within the mean-field theory (MFT) with the no-sea approximation~\cite{cw1}, 
quantum hadrodynamics (QHD) has described the bulk properties not only of 
infinite matter but of finite spherical, deformed and rotating nuclei 
successfully~\cite{rev1,rev2}.  These successes indicate that the 
particle-hole (p-h) interaction in QHD is realistic. However, various 
observables of nuclear many-body systems are sensitive to the single-particle 
properties around the Fermi surface. The knowledge of the residual 
particle-particle (p-p) interaction is required to describe them. 
Since this is still less understood, non-relativistic interactions such as 
Gogny force are used in the pairing channel in practical ``relativistic" MFT 
calculations. 

 The first study of a relativistic p-p interaction for the pairing channel in 
the nuclear medium was done by Kucharek and Ring~\cite{kr}. They adopted, as 
the particle-particle interaction ($v_{\rm pp}$) in the gap equation, a 
one-boson-exchange (OBE) interaction with the ordinary relativistic MFT 
parameters which gave the saturation under the no-sea approximation. 
The resulting maximum gap was about three times larger than the accepted 
values in the non-relativistic 
calculations~\cite{nrel0,nrel1,baldo,nrel2,nrel3,nrel4}. Various 
modifications to improve this result were proposed. They can be classified 
into two groups: One is to adopt the $v_{\rm pp}$ which is consistent with the 
p-h channel~\cite{rel1,rel2,mr}, and the other is to adopt effective ones 
which are not explicitly consistent with the p-h channel~\cite{rr,rev1,mt}. 
Among the former, we examined the p-h polarization in $v_{\rm pp}$ which 
reduced the pairing gap in the non-relativistic models~\cite{pol1,pol2}. 
But the result was negative; this suggests that the nucleon-antinucleon 
(N-$\bar{\rm N}$) polarization should be taken into account 
simultaneously~\cite{mr}. Before doing this, the antinucleon degrees of 
freedom have to be taken into account in the OBE step as did by Guimar\~aes 
{\it et al}~\cite{rel1}. In this paper, we discuss a more important effect of 
the Dirac sea than that they discussed.

 Our formulation consists of three steps. In the first step, the equations 
of motion of the normal and the anomalous Green's functions, $G$ and $F$, 
are derived. We start from the ordinary $\sigma$-$\omega$ model Lagrangian 
density,
\begin{eqnarray}
{\cal L}&=&\bar\psi(i\gamma_\mu\partial^\mu-M)\psi
\nonumber\\
 &+&{1\over2}(\partial_\mu\sigma)(\partial^\mu\sigma)
  -{1\over2}m_\sigma^2\sigma^2
  -{1\over4}\Omega_{\mu\nu}\Omega^{\mu\nu}
  +{1\over2}m_\omega^2\omega_\mu\omega^\mu
  +g_\sigma\bar\psi\sigma\psi-g_\omega\bar\psi\gamma_\mu\omega^\mu\psi ,
\nonumber\\
&\Omega_{\mu\nu}&=\partial_\mu\omega_\nu-\partial_\nu\omega_\mu .
\label{one}
\end{eqnarray}
\noindent
The Hamiltonian density ${\cal H}$ is derived from ${\cal L}$ and the 
chemical potential $\mu$ is introduced in the Hamiltonian, 
$H'=\int({\cal H}-\mu\psi^\dagger\psi)d^3x$.
The equations of motion for 
\begin{eqnarray}
 & &G_{\alpha\beta}(x-x')
=-i\langle\tilde0\vert T\psi_\alpha(x)\bar\psi_\beta(x')\vert\tilde0\rangle ,
\nonumber \\
 & &F_{\alpha\beta}(x-x')
=-i\langle\tilde0\vert T\bar\psi_\alpha(x)\bar\psi_\beta(x')\vert
   \tilde0\rangle ,
\label{two}
\end{eqnarray}
\noindent
are derived from $i\partial_t\psi=[\psi,H']$. Here we note that the Green's 
functions have to be defined by the superfluid ground state 
$\vert\tilde0\rangle$ in order to introduce the antinucleon in the next step. 
Aside from this, this first step is essentially the same as the formulation 
of Ref.\cite{kr}. After evaluating the commutator with the interaction term 
in $H'$, the meson fields are eliminated by using the inverse of the 
Klein-Gordon equations. Here the gauge term in the propagator of the $\omega$ 
meson can be discarded since $\omega$ couples to the baryon current which is 
conserved in average in the present superfluid case. The result is
\begin{eqnarray}
 & &(\Slash{p}-M+\gamma^0\mu)_{\alpha\gamma}G_{\gamma\beta}(x-x')
    =\delta_{\alpha\beta}\delta^4(x-x')
    +iV_{\alpha\gamma,\delta\epsilon}
     \langle\tilde0\vert T\psi_\epsilon(t,{\bf y})\bar\psi_\delta(t,{\bf y})
                          \psi_\gamma(x)\bar\psi_\beta(x')\vert\tilde0\rangle ,
\nonumber \\
 & &(-\Slash{\tilde p}-M+\gamma^0\mu)_{\gamma\alpha}F_{\gamma\beta}(x-x')
    =iV_{\gamma\alpha,\delta\epsilon}
     \langle\tilde0\vert T\psi_\epsilon(t,{\bf y})\bar\psi_\delta(t,{\bf y})
                     \bar\psi_\gamma(x)\bar\psi_\beta(x')\vert\tilde0\rangle ,
\nonumber \\
 & & V_{\alpha\gamma,\delta\epsilon}
 =-g_\sigma^2D_\sigma\delta_{\alpha\gamma}\delta_{\delta\epsilon}
  +g_\omega^2D_\omega(\gamma^\mu)_{\alpha\gamma}(\gamma_\mu)_{\delta\epsilon} ,
\nonumber \\
 & & D_i={1\over{-\triangle+m_i^2}} , \quad (i=\sigma,~\omega) ,
\label{three}
\end{eqnarray}
\noindent
with $x=(t,{\bf x})$ and 
$\Slash{\tilde p}=i\beta(\partial_t-{\bf\alpha}\cdot{\bf\nabla})$. 
Here we introduced an instantaneous approximation since it was reported in the 
preceeding works~\cite{rel1,rel2} that the retardation effects are small. 
The time-ordered products in the right-hand side are decomposed \`a la 
Gor'kov~\cite{gor} by extending Wick's theorem,
\begin{eqnarray}
 & &\langle\tilde0\vert T\psi_\epsilon(t,{\bf y})\bar\psi_\delta(t,{\bf y})
                          \psi_\gamma(x)\bar\psi_\beta(x')\vert\tilde0\rangle
\nonumber \\
 & &=-\rho_{\epsilon\delta'}(0)\gamma^0_{\delta'\delta}
           iG_{\gamma\beta}(x-x')
     +\rho_{\gamma\delta'}({\bf x-y})\gamma^0_{\delta'\delta}
           iG_{\epsilon\beta}(y-x')
     -\kappa_{\gamma\epsilon}({\bf x-y})iF_{\delta\beta}(y-x') ,
\nonumber \\
 & &\langle\tilde0\vert T\psi_\epsilon(t,{\bf y})\bar\psi_\delta(t,{\bf y})
                     \bar\psi_\gamma(x)\bar\psi_\beta(x')\vert\tilde0\rangle
\nonumber \\
 & &=-\rho_{\epsilon\delta'}(0)\gamma^0_{\delta'\delta}
           iF_{\gamma\beta}(x-x')
     +\rho_{\epsilon\gamma'}({\bf y-x})\gamma^0_{\gamma'\gamma}
          iF_{\delta\beta}(y-x')
\nonumber \\
 & & +\kappa_{\delta'\gamma'}^\ast({\bf y-x})\gamma^0_{\delta'\delta}
          \gamma^0_{\gamma'\gamma}iG_{\epsilon\beta}(y-x') ,
\label{four}
\end{eqnarray}
\noindent
where the normal and the anomalous densities are 
\begin{eqnarray}
 & &\rho_{\alpha\beta}({\bf x-y})
 =\langle\tilde0\vert \psi^\dagger_\beta(t,{\bf y})
                      \psi_\alpha(x)\vert\tilde0\rangle ,
\nonumber \\
 & &\kappa_{\alpha\beta}({\bf x-y})
 =\langle\tilde0\vert \psi_\beta(t,{\bf y})
                      \psi_\alpha(x)\vert\tilde0\rangle .
\label{five}
\end{eqnarray}
\noindent
The selfenergy and the pairing field are defined by these densities  as
\begin{eqnarray}
 & &\Sigma_{\alpha\gamma}({\bf x-y})
  =(V_{\alpha\gamma,\delta\epsilon}\rho_{\epsilon\delta'}(0)
   -V_{\alpha\epsilon,\delta\gamma}\rho_{\epsilon\delta'}({\bf x-y}))
   \gamma^0_{\delta'\delta} ,
\nonumber \\
& &\Delta_{\alpha\gamma}({\bf x-y})
 =\gamma^0_{\alpha\alpha'}\gamma^0_{\gamma\gamma'}
   V_{\alpha'\delta,\gamma'\epsilon}\kappa_{\delta\epsilon}({\bf x-y}) .
\label{six}
\end{eqnarray}
\noindent
The Fourier transform of the equations of motion for $g$ and $f$ defined by 
$G_{\gamma\beta}=g_{\gamma\beta'}\gamma^0_{\beta'\beta}$ and 
$F_{\gamma\beta}=f_{\gamma'\beta'}
\gamma^0_{\gamma'\gamma}\gamma^0_{\beta'\beta}$ 
is given by
\begin{equation}
\left(
\begin{array}{@{\,}cc@{\,}}
 (\omega-h+\mu)_{\alpha\gamma} & -\Delta_{\alpha\gamma} \\
 \Delta_{\alpha\gamma}^\ast & (\omega+h^\ast-\mu)_{\alpha\gamma}
\end{array}
\right)
\left(
\begin{array}{@{\,}c@{\,}}
 g_{\gamma\beta} \\
 f_{\gamma\beta}
\end{array}
\right)
=
\left(
\begin{array}{@{\,}c@{\,}}
 \delta_{\alpha\beta} \\
 0
\end{array}
\right) ,
\label{seven}
\end{equation}
\noindent
where $h={\bf\alpha}\cdot{\bf k}+\beta(M+\Sigma(k))$.

 In the second step, we derive the equations for the Bogoliubov transformation 
amplitudes by expressing $g$ and $f$ in Eq.(\ref{seven}) in terms of them. 
We introduce the Dirac field in the Schr\"odinger picture,
\begin{equation}
\psi({\bf x})=\frac{1}{\sqrt{V}}\sum_{\lambda=({\bf k}s)}
    (a_\lambda U(\lambda){\rm e}^{i{\bf kx}}
    +b_\lambda^\dagger V(\lambda){\rm e}^{-i{\bf kx}}) ,
\label{eight}
\end{equation}
\noindent
to manipulate the $t$-dependence. Here the normalization of $U$ and $V$ is 
chosen conforming to Ref.\cite{sw}. The Bogoliubov amplitudes are defined by
\begin{eqnarray}
 & &A_\lambda=\langle\tilde0\vert a_\lambda\eta_\lambda^\dagger
                            \vert\tilde0\rangle ,\quad 
    B_\lambda=\langle\tilde0\vert b_{-\lambda}^\dagger\eta_\lambda^\dagger
                            \vert\tilde0\rangle ,
\nonumber \\
 & &C_\lambda=\langle\tilde0\vert a_{-\lambda}^\dagger\eta_\lambda^\dagger
                            \vert\tilde0\rangle ,\quad 
    D_\lambda=\langle\tilde0\vert b_\lambda\eta_\lambda^\dagger
                            \vert\tilde0\rangle ,
\label{nine}
\end{eqnarray}
\noindent
as the overlaps between the quasiparticle $\eta_\lambda^\dagger$,
\begin{equation}
 H'\eta_\lambda^\dagger\vert\tilde0\rangle
 ={\cal E}_k\eta_\lambda^\dagger\vert\tilde0\rangle , \quad
 H'\vert\tilde0\rangle=0 ,
\label{ten}
\end{equation}
\noindent
and the nucleon or the antinucleon. These give the explicit expressions of 
$g$ and $f$, 
\begin{eqnarray}
g_{\gamma\beta}(\omega,{\bf k})=\frac{1}{\omega-{\cal E}_k+i\epsilon}
\sum_{s}
& &\bigl\{
     \vert A_{\lambda}\vert^2U_\gamma(\lambda)U_\beta^\ast(\lambda)
    +\vert B_{\lambda}\vert^2V_\gamma(-\lambda)V_\beta^\ast(-\lambda)
   \bigr.
\nonumber \\
& &\bigl.
    +A_{\lambda}B_{\lambda}^\ast U_\gamma(\lambda)V_\beta^\ast(-\lambda)
    +B_{\lambda}A_{\lambda}^\ast V_\gamma(-\lambda)U_\beta^\ast(\lambda)
   \bigr\}
\nonumber \\
+\frac{1}{\omega+{\cal E}_k-i\epsilon}
\sum_{s}
& &\bigl\{
     \vert C_{-\lambda}\vert^2U_\gamma(\lambda)U_\beta^\ast(\lambda)
    +\vert D_{-\lambda}\vert^2V_\gamma(-\lambda)V_\beta^\ast(-\lambda)
   \bigr.
\nonumber \\
& &\bigl.
    +D_{-\lambda}C_{-\lambda}^\ast U_\gamma(\lambda)V_\beta^\ast(-\lambda)
    +C_{-\lambda}D_{-\lambda}^\ast V_\gamma(-\lambda)U_\beta^\ast(\lambda)
   \bigr\} ,
\nonumber \\
f_{\gamma\beta}(\omega,{\bf k})=\frac{1}{\omega-{\cal E}_k+i\epsilon}
\sum_{s}
& &\bigl\{
     C_{\lambda}A_{\lambda}^\ast U_\gamma^\ast(-\lambda)U_\beta^\ast(\lambda)
    +D_{\lambda}B_{\lambda}^\ast V_\gamma^\ast(\lambda)V_\beta^\ast(-\lambda)
   \bigr.
\nonumber \\
& &\bigl.
    +C_{\lambda}B_{\lambda}^\ast U_\gamma^\ast(-\lambda)V_\beta^\ast(-\lambda)
    +D_{\lambda}A_{\lambda}^\ast V_\gamma^\ast(\lambda)U_\beta^\ast(\lambda)
   \bigr\}
\nonumber \\
+\frac{1}{\omega+{\cal E}_k-i\epsilon}
\sum_{s}
& &\bigl\{
     C_{-\lambda}A_{-\lambda}^\ast U_\gamma^\ast(-\lambda)U_\beta^\ast(\lambda)
    +D_{-\lambda}B_{-\lambda}^\ast V_\gamma^\ast(\lambda)V_\beta^\ast(-\lambda)
   \bigr.
\nonumber \\
& &\bigl.
 +D_{-\lambda}A_{-\lambda}^\ast U_\gamma^\ast(-\lambda)
                                   V_\beta^\ast(-\lambda)
    +C_{-\lambda}B_{-\lambda}^\ast V_\gamma^\ast(\lambda)U_\beta^\ast(\lambda)
   \bigr\} ,
\label{eleven}
\end{eqnarray}
\noindent
the former describes the normal propagation and the $\Delta N=0$ pairing, and 
the latter describes the ordinary $\Delta N=2$ pairing. This $\Delta N=0$ 
pairing appears because Cooper pairs can be formed regardless of the sign of 
the single-nucleon energy. Substituting Eq.(\ref{eleven}) into 
Eq.(\ref{seven}) and defining the matrix elements of the pairing field 
$\Delta$ as
\begin{eqnarray}
& &\Delta(\lambda)
   ={\rm e}^{-i\alpha_+(\lambda)}U^\dagger_\alpha(\lambda)
   \Delta_{\alpha\gamma}T_{\gamma\beta}U_\beta(\lambda) ,
\nonumber \\
& &\tilde\Delta(\lambda)
   ={\rm e}^{-i\alpha_-(-\lambda)}V^\dagger_\alpha(-\lambda)
    \Delta_{\alpha\gamma}T_{\gamma\beta}V_\beta(-\lambda) ,
\nonumber \\
& &\delta(\lambda)
   ={\rm e}^{-i\alpha_-(-\lambda)}U^\dagger_\alpha(\lambda)
    \Delta_{\alpha\gamma}T_{\gamma\beta}V_\beta(-\lambda) ,
\label{twelve}
\end{eqnarray}
\noindent
for the Fermi sea pairing, the Dirac sea pairing, and the $\Delta N=0$ 
pairing, respectively, we obtain the equation of the Bogoliubov amplitudes, 
\begin{equation}
\left(
\begin{array}{@{\,}cccc@{\,}}
 \omega-E_k+\mu & 0 & -\Delta(\lambda) & -\delta(\lambda) \\
 0 & \omega+E_k+\mu & -\delta(\lambda) & -\tilde\Delta(\lambda) \\
 -\Delta(\lambda) & -\delta(\lambda) & \omega+E_k-\mu & 0 \\
 -\delta(\lambda) & -\tilde\Delta(\lambda) & 0 & \omega-E_k-\mu
\end{array}
\right)
\left(
\begin{array}{@{\,}c@{\,}}
 A_{\lambda} \\
 B_{\lambda} \\
 C_{\lambda} \\
 D_{\lambda} \\
\end{array}
\right)
=0 .
\label{thirt}
\end{equation}
\noindent
Here we used
\begin{eqnarray}
 & &h_{\alpha\gamma}U_\gamma(\lambda)=E_kU_\alpha(\lambda) ,\quad
    h_{\alpha\gamma}V_\gamma(-\lambda)=-E_kV_\alpha(-\lambda) ,
\nonumber \\
 & &E_k=\sqrt{{\bf k}^2+M^{\star2}} ,
\nonumber \\
 & &TU(\lambda)={\rm e}^{i\alpha_+(\lambda)}U^\ast(-\lambda) ,\quad
    TV(-\lambda)={\rm e}^{i\alpha_-(-\lambda)}V^\ast(\lambda) ,
\nonumber \\
 & &Th^\ast=hT , \quad T\Delta^\ast=\Delta T , \quad T=i\gamma^1\gamma^3 .
\label{fort}
\end{eqnarray}
\noindent
Note that we adopted notations such that Eq.(\ref{thirt}) took the same form 
as that of Ref.\cite{rel1} and chose a phase convention such that all the 
matrix elements were real. Among the eigenvalues,
\begin{equation}
 \omega^2=\bigl(E_k^2+\mu^2+\frac{1}{2}(\Delta^2+\tilde\Delta^2)+\delta^2\bigr)
 -\frac{1}{2}\sqrt{(4E_k\mu-\Delta^2+\tilde\Delta^2)^2
                   +4\delta^2\bigl(4E_k^2+(\Delta+\tilde\Delta)^2\bigr)}
\label{fift}
\end{equation}
\noindent
corresponds to the Fermi sea pairing in the decoupling ($\delta\rightarrow0$) 
limit.

 In the third step, we express $\kappa$, and subsequently the matrix elements 
$\Delta(\lambda)$, $\tilde\Delta(\lambda)$, and $\delta(\lambda)$, 
in terms of $A_\lambda$ - $D_\lambda$. The Fourier transform of 
$\kappa$ in Eq.(\ref{five}) is given by
\begin{eqnarray}
\kappa_{\delta\epsilon}(\omega,{\bf k})
   =2\pi\delta(\omega+{\cal E}_k)\sum_{s}
& &\bigl\{
   A_{-\lambda}C_{-\lambda}U_\epsilon(-\lambda)U_\delta(\lambda)
  +B_{-\lambda}D_{-\lambda}V_\epsilon(\lambda)V_\delta(-\lambda)
   \bigr.
\nonumber \\
& &\bigl.
  +A_{-\lambda}D_{-\lambda}U_\epsilon(-\lambda)V_\delta(-\lambda)
  +B_{-\lambda}C_{-\lambda}V_\epsilon(\lambda)U_\delta(\lambda)
\bigr\} .
\label{sixt}
\end{eqnarray}
\noindent
Substituting this into the Fourier transform of the second equation of 
(\ref{six}) gives 
\begin{eqnarray}
& &\Delta(\lambda)
    =-\frac{i}{2}\int\frac{d^3p}{(2\pi)^3}\frac{1}{4E_kE_p}
\Bigl(\frac{-g_\sigma^2}{\vert{\bf k}-{\bf p}\vert^2+m_\sigma^2}
\Bigr.
\nonumber \\
&\bigl[&{\rm Tr}\bigl\{(\Slash{k}+M^\star)(\Slash{p}+M^\star)\bigr\} 
      {\rm e}^{i\alpha_+(\lambda)}A_{\lambda}C_{\lambda}
       +{\rm Tr}\bigl\{(\Slash{k}+M^\star)(\Slash{\tilde p}-M^\star)\bigr\} 
      {\rm e}^{i\alpha_-(-\lambda)}B_{\lambda}D_{\lambda}
\bigr]
\nonumber \\
& &\quad\quad\quad
     +\frac{g_\omega^2}{\vert{\bf k}-{\bf p}\vert^2+m_\omega^2}
\nonumber \\
&\bigl[&{\rm Tr}\bigl\{(\Slash{k}+M^\star)\gamma_\mu
                       (\Slash{p}+M^\star)\gamma_\mu\bigr\} 
      {\rm e}^{i\alpha_+(\lambda)}A_{\lambda}C_{\lambda}
\Bigl.
       +{\rm Tr}\bigl\{(\Slash{k}+M^\star)\gamma_\mu
                     (\Slash{\tilde p}-M^\star)\gamma_\mu\bigr\} 
      {\rm e}^{i\alpha_-(-\lambda)}B_{\lambda}D_{\lambda}
\bigr] \Bigr),
\nonumber \\
& &\tilde\Delta(\lambda)
    =-\frac{i}{2}\int\frac{d^3p}{(2\pi)^3}\frac{1}{4E_kE_p}
\Bigr(\frac{-g_\sigma^2}{\vert{\bf k}-{\bf p}\vert^2+m_\sigma^2}
\Bigl.
\nonumber \\
&\bigl[&{\rm Tr}\bigl\{(\Slash{\tilde k}-M^\star)(\Slash{p}+M^\star)
                \bigr\} 
      {\rm e}^{i\alpha_+(\lambda)}A_{\lambda}C_{\lambda}
       +{\rm Tr}\bigl\{(\Slash{\tilde k}-M^\star)(\Slash{\tilde p}-M^\star)
              \bigr\} 
      {\rm e}^{i\alpha_-(-\lambda)}B_{\lambda}D_{\lambda}
\bigr]
\nonumber \\
& &\quad\quad\quad
     +\frac{g_\omega^2}{\vert{\bf k}-{\bf p}\vert^2+m_\omega^2}
\nonumber \\
&\bigl[&{\rm Tr}\bigl\{(\Slash{\tilde k}-M^\star)\gamma_\mu
                       (\Slash{p}+M^\star)\gamma_\mu\bigr\} 
      {\rm e}^{i\alpha_+(\lambda)}A_{\lambda}C_{\lambda}
\Bigl.
       +{\rm Tr}\bigl\{(\Slash{\tilde k}-M^\star)\gamma_\mu
                     (\Slash{\tilde p}-M^\star)\gamma_\mu\bigr\} 
      {\rm e}^{i\alpha_-(-\lambda)}B_{\lambda}D_{\lambda}
\bigr] \Bigr),
\nonumber \\
& &\delta(\lambda)
    =-\frac{i}{2}\int\frac{d^3p}{(2\pi)^3}\frac{1}{4E_kE_p}
\Bigr(\frac{-g_\sigma^2}{\vert{\bf k}-{\bf p}\vert^2+m_\sigma^2}
\Bigl.
\nonumber \\
&\bigl[&{\rm Tr}\bigl\{\gamma^5(\Slash{k}-M^\star)\gamma^0
                       \gamma^5(\Slash{p}-M^\star)\gamma^0
                \bigr\} 
      {\rm e}^{i\alpha_+(\lambda)}B_{\lambda}C_{\lambda}
\bigr.
\nonumber \\
&\bigl.&
       -{\rm Tr}\bigl\{\gamma^5(\Slash{k}-M^\star)\gamma^0
                     \gamma^5(\Slash{\tilde p}+M^\star)\gamma^0
              \bigr\} 
      {\rm e}^{i\alpha_-(-\lambda)}A_{\lambda}D_{\lambda}
\bigr] 
\nonumber \\
& &\quad\quad\quad
     +\frac{g_\omega^2}{\vert{\bf k}-{\bf p}\vert^2+m_\omega^2}
\nonumber \\
&\bigl[&{\rm Tr}\bigl\{\gamma^5(\Slash{k}-M^\star)\gamma^0\gamma_\mu
                       \gamma^5(\Slash{p}-M^\star)\gamma^0\gamma_\mu
                \bigr\} 
      {\rm e}^{i\alpha_+(\lambda)}B_{\lambda}C_{\lambda}
 \bigr.
\nonumber \\
\Bigl.
&\bigl.&
       -{\rm Tr}\bigl\{\gamma^5(\Slash{k}-M^\star)\gamma^0\gamma_\mu
                     \gamma^5(\Slash{\tilde p}+M^\star)\gamma^0\gamma_\mu
              \bigr\} {\rm e}^{i\alpha_-(-\lambda)}A_{\lambda}D_{\lambda}
\bigr] \Bigr),
\label{sevent}
\end{eqnarray}
\noindent
where both $\gamma_\mu$ in each line are covariant due to the time reversal. 
Here we used $A_{-\lambda}D_{-\lambda}=-B_{\lambda}C_{\lambda}$, and so on, 
derived from the definition of $\kappa$, the $s$-independence of 
${\rm e}^{i\alpha_+(\lambda)}A_{\lambda}C_{\lambda}$, and so on, and 
\begin{eqnarray}
& &\sum_sU(\lambda)\bar U(\lambda)
    =\frac{1}{2E_k}(\Slash{k}+M^\star) ,\quad
   \sum_sV(-\lambda)\bar V(-\lambda)
    =\frac{1}{2E_k}(\Slash{\tilde k}-M^\star) ,
\nonumber \\
& &\sum_sU(\lambda)\bar V(-\lambda)
    =-\frac{1}{2E_k}\gamma^5(\Slash{k}-M^\star)\gamma^0 ,\quad
   \sum_sV(-\lambda)\bar U(\lambda)
    =\frac{1}{2E_k}\gamma^5(\Slash{\tilde k}+M^\star)\gamma^0 ,
\nonumber \\
& &\tilde k=(E_k,-{\bf k}) .
\label{eighte}
\end{eqnarray}
\noindent
Equations (\ref{thirt}) and (\ref{sevent}) form a system of selfconsistent 
equations. The nucleon effective mass $M^\star$ is determined by the 
selfconsistent condition for the scalar part of the selfenergy, the first 
equation of (\ref{six}), 
\begin{eqnarray}
  M^\star&=&M-{{g_\sigma^2}\over{m_\sigma^2}}{\gamma\over{2\pi^2}}
 \int_0^{\Lambda_{\rm c}} {M^\star\over\sqrt{{\bf p}^2+M^{\star\,2}}}C_p^2p^2dp
\nonumber\\
 &+&{{g_\sigma^2}\over{m_\sigma^2}}{1\over{\pi^2}}
    \bigl\{M^{\star3}\ln{\bigl({{M^{\star}}\over M}\bigr)}-M^2(M^\star-M)
   -{5\over2}M(M^\star-M)^2-{11\over6}(M^\star-M)^3\bigr\} ,
\label{ninet}
\end{eqnarray}
\noindent
within the relativistic Hartree approximation (RHA) in which the divergence 
due to the Dirac sea is renormarized using the counter terms
\begin{equation}
 {\cal L}_{\rm ct}=\alpha_1\sigma+\frac{1}{2!}\alpha_2\sigma^2
  +\frac{1}{3!}\alpha_3\sigma^3+\frac{1}{4!}\alpha_4\sigma^4 .
\label{twenty}
\end{equation}
\noindent
Hereafter we suppress the argument $s$, and therefore the direction of 
${\bf k}$ which defines the direction of $s$, of the gaps and the Bogoliubov 
amplitudes because they affect only the overall sign. The expression of the 
vacuum contribution in Eq.(\ref{ninet}) is taken from the non-superfluid 
case~\cite{sw}. We will discuss this later. Then the actual task is to solve 
the coupled equations (\ref{thirt}), (\ref{sevent}), and (\ref{ninet}). 
If we neglect the vacuum fluctuation contribution in Eq.(\ref{ninet}), the 
system of equations corresponds to that of Ref.\cite{rel1}. 
Here we note that we adopt the Hartree approximation with or without the 
Dirac sea contribution as in Ref.\cite{kr} for the no-sea case.

 Now we proceed to the numerical results. Parameters used are $M=~$939 MeV, 
$m_\sigma=~$550 MeV, $m_\omega=~$783 MeV, $g_\sigma^2=~$62.89 for the RHA or 
91.64 for the MFT, $g_\omega^2=~$79.78 for the RHA or 136.2 for the 
MFT~\cite{sw}, and $\gamma=~$4 (symmetric nuclear matter). 
First we look into the relative magnitudes of the gaps $\Delta(k)$, 
$\tilde\Delta(k)$, and $\delta(k)$. Since $A_k$ and $C_k$ are dominant among 
those associated with the eigenvalue Eq.(\ref{fift}), $\Delta(k)$ and 
$\tilde\Delta(k)$ are much larger than $\delta(k)$. From the definition of 
$\kappa$ in Eq.(\ref{five}), two kinds of terms, one is antisymmetric in spin 
space and even with respect to the inversion of the momentum, and the other is 
symmetric and odd, are possible in $\kappa_{\delta\epsilon}(\omega,{\bf k})$. 
If the finite-range effects are neglected, only six terms of the former type 
are possible as discussed in Ref.\cite{rel2}. They correspond to scalar, 
pseudoscalar, and 4-vector terms in $\kappa T$. Among them, the scalar and the 
time component of the vector are dominant and consequently $\Delta(k)$ and 
$\tilde\Delta(k)$ are determined by these two terms as pointed out in 
Ref.\cite{rel2}. Basically this applies also to the calculations of 
Ref.\cite{rel1}~ and ours in which the terms of the latter type are also 
included. In contrast, $\delta(k)$ which measures the $\Delta N=0$ pairing is 
given by the pseudoscalar and pseudovector terms in $\kappa T$. Since they 
contain $B_kC_k$ and $A_kD_k$, their typical values are the order of 
$10^{-6}~$MeV or less both in the RHA and in the MFT. Therefore, as for the 
magnitude of the $\Delta N=0$ coupling effect on $\Delta(k)$, our calculation 
does not agree with that of Ref.\cite{rel1}. The origin of this disagreement 
is that their Eq.(54) does not have this structure that the different type of 
products of the Bogoliubov amplitudes appear in $\delta(k)$. The present 
result indicates that the Fermi sea pairing and the Dirac sea pairing decouple 
in accuracy of $10^{-7}$. Accordingly, we use the expressions for 
$\delta(k)=0$ hereafter. Since the equations for $A_k$ and $C_k$ decoupled 
from $B_k$ and $D_k$ are equivalent to a gap equation,
\begin{equation}
\Delta(k)=-{1\over{8\pi^2}}
  \int_0^{\Lambda_{\rm c}} \bar v_{\rm pp}(k,p)
         {\Delta(p)\over\sqrt{(E_p-E_{k_{\rm F}})^2+\Delta^2(p)}}p^2dp ,
\label{twentyone}
\end{equation}
\noindent
where $\bar v_{\rm pp}(k,p)$ is an antisymmetrized matrix element of the 
adopted p-p interaction with an instantaneous approximation and an integration 
with respect to the angle between ${\bf k}$ and ${\bf p}$ to project out the 
$S$-wave component, the numerical task is greatly simpified to solving the 
coupled equations (\ref{ninet}) and (\ref{twentyone}). 

\begin{figure}[t]
\begin{center}
\epsfig{figure=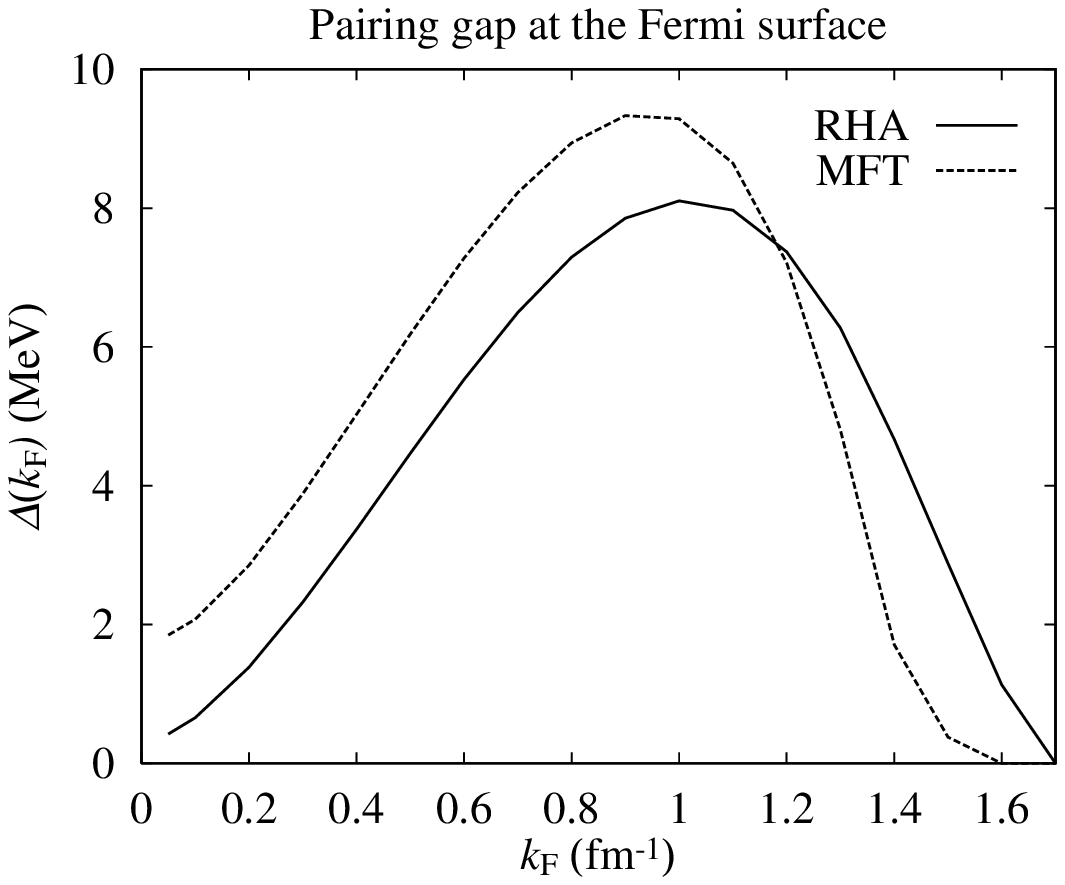,width=12cm}
\end{center}
\caption{
Pairing gap in symmetric nuclear matter at the Fermi surface as functions of 
the Fermi momentum. Solid and dashed lines indicate the results obtained by 
taking and not taking into account the vacuum fluctuation contribution, 
respectively.
}
\label{figa}
\end{figure}

\begin{figure}[h]
\begin{center}
\epsfig{figure=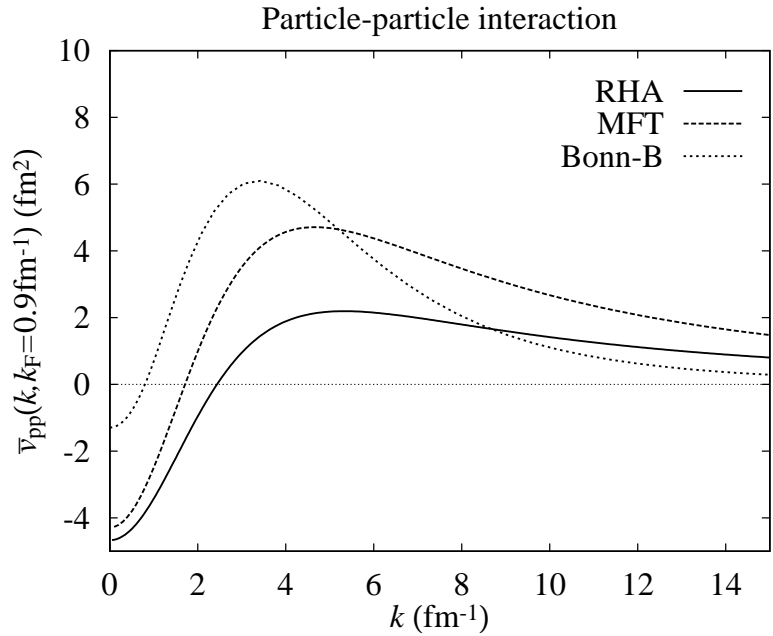,width=12cm}
\end{center}
\caption{
Matrix element $\bar v_{\rm pp}(k,k_{\rm F})$ as functions of the 
momentum $k$, with a Fermi momentum $k_{\rm F}=~$0.9 fm$^{-1}$. Solid and 
dashed lines indicate the results obtained by taking and not taking into 
account the vacuum fluctuation contribution, respectively. The Bonn-B 
potential for the $^1S_0$ channel is also shown by the dotted line. 
}
\label{figb}
\end{figure}

\begin{figure}[h]
\begin{center}
\epsfig{figure=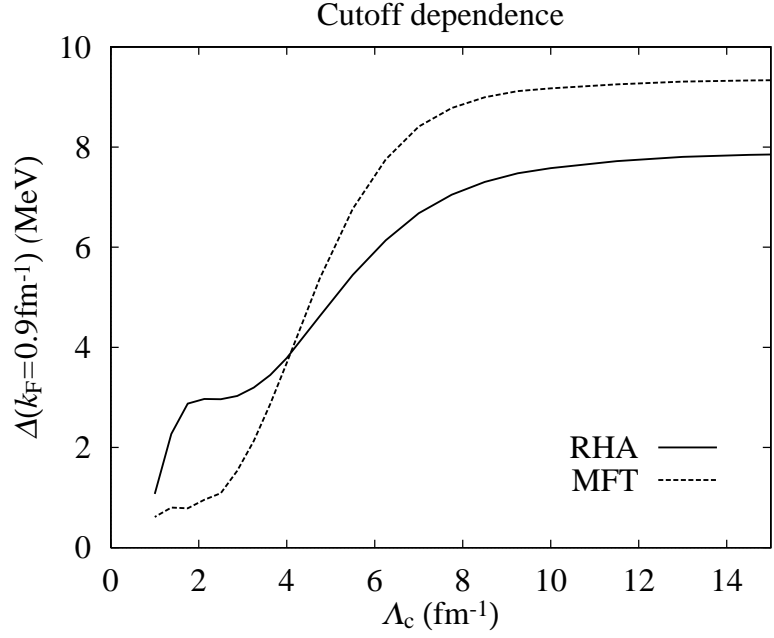,width=12cm}
\end{center}
\caption{
Pairing gap at the Fermi surface, $k_{\rm F}=~$0.9 fm$^{-1}$, as functions of 
the cutoff parameter in the numerical integrations. Solid and 
dashed lines indicate the results obtained by taking and not taking into 
account the vacuum fluctuation contribution, respectively. 
}
\label{figc}
\end{figure}

 Next we compare the results of the RHA and the MFT obtained by adopting 
$\Lambda_{\rm c}=~$15 fm$^{-1}$ which is large enough for the numerical 
integrations to converge (see Fig.\ref{figc}). These are presented in 
Fig.\ref{figa}. This shows that the Dirac sea contributes to reducing the gap 
at low density while to enhancing it at high density. This can be understood 
as follows: Although both $g_\sigma$ and $g_\omega$ are reduced in the RHA in 
comparison with the MFT, altogether they act to reduce the repulsion as shown 
in Fig.\ref{figb}. Since both the low-momentum attraction and the 
high-momentum repulsion give positive contributions to $\Delta(k_{\rm F})$ as 
discussed in Refs.\cite{rr,mt}, the above-mentioned reduction of the repulsion 
leads to reduction of $\Delta(k_{\rm F})$ in the RHA. Besides, the 
low-momentum attraction decreases steeply as density increases especially in 
the MFT (not shown). This leads to steeper reduction of $\Delta(k_{\rm F})$ at 
high density in the MFT than in the RHA. Here we note that this RHA 
calculation was done by assuming the same vacuum fluctuation contribution as 
the non-superfluid case as mentioned above. We believe that this approximation 
is practical because the effect of the pairing on $M^\star$ is negligible 
numerically except at very low density. This indicates that the bulk property, 
$M^\star$, affects the Fermi-surface property, superfluidity, whereas the 
opposite is not true. 

 Since QHD is an effective theory of hadrons, form factors or cutoffs related 
to the spatial size of hadrons might be necessary~\cite{saga}. On the other 
hand, an important feature of the gap equation is that it has such a form that 
the short range correlation is involved~\cite{cor1,cor2,nrel2}. This leads to 
a similar oscillatory behavior of $\Delta(k)$ to that of 
$\bar v_{\rm pp}(k,k_{\rm F})$ as functions of $k$~\cite{baldo,rr,mt}. 
Matera {\it et al}. found that their formulation for the pairing gap gave a 
cutoff~\cite{rel2}. This is interesting in respect that both which are 
originated from the pairing correlation act to evade the repulsion at high 
momentum. They reported that the values they obtained were 1.7 - 1.9 
fm$^{-1}$. These values mean that the repulsive part was completely cut, see 
Fig.\ref{figb}. Since our formulation does not have any means to choose a 
cutoff as those of Refs.\cite{kr,rel1}, here we show the dependence of 
$\Delta(k_{\rm F})$ at $k_{\rm F}=~$0.9 fm $^{-1}$, where it becomes maximum, 
on the cutoff momentum $\Lambda_{\rm c}$ as a free parameter. The result is 
presented in Fig.\ref{figc}. This shows that the MFT result decreases steeply 
as the cutoff decreases. This can be understood as follows: The difference 
between the RHA and the MFT in the low-density case in Fig.\ref{figa}~ mainly 
comes from the difference in the magnitude of the high-momentum repulsion. 
In other words, the contribution of the high-momentum repulsion is more 
important in the MFT case. Consequently, as the cutoff momentum decreases, 
the $\Delta(k_{\rm F})$ of the MFT calculation decreases more. 
The plateaus around $\Lambda_{\rm c}=~$2 - 3 fm$^{-1}$ in the RHA and 
$\Lambda_{\rm c}=~$1 - 2 fm$^{-1}$ in the MFT are due to the sign change of
$v_{\rm pp}$ around there (see Fig.\ref{figb}). A comparison with the Bonn-B 
potential~\cite{bonn} which reproduced the maximum pairing gap accepted in 
the non-relativistic studies~\cite{rr,rev1} is also shown in Fig.\ref{figb}. 
This indicates that the high-momentum repulsion of the MFT without a cutoff 
is too strong and the low-momentum repulsion is too weak both in the RHA and 
in the MFT to give reasonable pairing gaps. A possible improvement in the 
p-p channel would be to take into account the nucleon-antinucleon 
(N-$\bar{\rm N}$) polarization. 

 To summarize, we have studied two kinds of Dirac sea effects on the pairing 
gap in nuclear matter. One is the vacuum fluctuation effect on the nucleon 
effective mass and the other is the direct coupling between the Fermi sea 
pairing and the Dirac sea pairing. The former is a bulk effect while the 
latter affects only the Fermi surface. Our calculation indicates that the 
former is more important. The dependence of $\Delta(k_{\rm F})$ on the 
high-momentum cutoff was also discussed and the MFT result has been shown 
to depend more strongly on it than the RHA one.

\vskip 0.5cm
 Discussions with Prof. M. Nakano and Prof. R. Tamagaki are acknowledged. 
The author also thanks Mr. T. Tanigawa for calculating the Bonn potential. 
Numerical calculations were done using the computer 
system of the Information Processing Center, Fukuoka University of Education.

\end{document}